\documentstyle[times,pramana,epsf,floats]{ias}
\begin{document}
\mark{{Quadrapole deformation }{P.A. Ganai et al }}
\title{Temperature and angular momentum dependence of the quadrupole deformation in sd-shell}

\author{P.A. Ganai$^{1}$, J.A. Sheikh$^{1}$, I. Maqbool$^{1}$ and R.P. Singh$^{2}$}
\address{$^1$Department of Physics, University of Kashmir, Srinagar,
190 006, India }
\address{$^2$Inter-University Accelerator Center, New Delhi, 110 067, India}

\abstract{Temperature and angular momentum dependence of the quadrupole deformation is studied
in the middle of the sd-shell for $^{28}$Si and $^{27}$Si isotopes
using the spherical shell model approach. The shell model
calculations have been performed using the standard USD interaction
and the canonical partition function constructed from the
calculated eigen-solutions. It is shown that the extracted average quadrupole
moments show a transitional behavior as a function of temperature and the infered
transitional temperature is shown to vary with angular-momentum.
The quadrupole deformation of the individual eigen-states
is also analyzed.
}

\maketitle

The study of phase transition in quantum many-body systems
is one of frontier research topics in physics. Phase transitions
are observed both in macroscopic as well as in small quantum many-body systems.
In macroscopic systems, for instance a metallic superconductor,
the linear dimension of the system is quite large
and the transition from one phase to the other occurs at one point \cite{jbardeen}. For 
small systems, the fluctuations play a very central
role and the existence of the discontinuity in the heat capacity critically depends on the number of
constituent particles in the system. This has been demonstrated in small metallic
grains that discontinuity is observed with large number of electrons in the grain.
However, as the number of electrons in the grain approaches around 100, the discontinuity
or the peak structure in the heat capacity disappears \cite{muh72,b88,sliu,b93}.

Phase transitions have also been studied extensively in atomic nuclei using the
Hartree-Fock-Bogoliubov (HFB) method.  The HFB theory predicts phase transition as a 
function of rotational frequency (angular momentum) and temperature (excitation energy). 
The shape transition as a function of rotational frequency has been well studied in most of 
the regions of the nuclear chart. In particular, in the rare-earth region the examples of 
the shape transition are documented in the text books \cite{rs80,szy83}. Most of the
nuclei in this region are prolate deformed with the rotational axis perpendicular to the
symmetry axis at low spins and it is known that this shape changes to oblate
non-collective at higher angular momenta in many nuclei in this region. For instance, 
in the case $^{160}$Yb, the shape is prolate for spins up to 40 $\hbar$ and above this
spin value the shape becomes oblate non-collective.  The phase transition has also been
investigated using the finite temperature HFB approach \cite{good84,good94} and 
Landau theory \cite{alh89}. The main conclusion from these studies is that nuclei which
are deformed at low temperatures exhibit a shape transition to spherical shape as the
temperature of the system is raised. The critical temperature at which this shape 
transition occurs has a maximum value for isotopes between magic numbers and in 
rare-earth nuclei this maximum temperature is about 1.85 MeV \cite{alh89}.

It is known that the phase transition obtained in the above studies critically depends 
on the inclusion of the quantal and statistical fluctuations \cite{good84,ross91}. In the 
absence of the fluctuations, the phase transition is sharp first order and when some 
aspects of the fluctuations are considered, the phase transition is smoothed out. These 
studies have been mostly performed for heavier nuclei using the grand canonical ensemble. 
The situation is very different for lighter mass nuclei as it is possible to perform the
exact shell model calculations and the canonical partition function can be constructed. 
The canonical partition function will incorporate fluctuations accurately. 
The purpose of the present work is to investigate the shape transition in the sd-shell 
using the canonical partition function constructed from the spherical shell model eigen-states. In a 
recent work, we have used this approach to study the pairing correlations \cite{she2008}.

The spherical shell model calculations have been performed for $^{28}$Si and $^{27}$Si
using the recently developed shell model program \cite{she2008,ssunp}.
The shell model Hamiltonian, generally, contains single-particle 
and two-body parts and in the second quantized notation is written as
\begin{equation}
\hat H\,=\,\hat h_{sp}\,+\,\hat V_{2},  \label{E01}
\end{equation}
where, 
\begin{equation}
\hat h_{sp}\,=\,\sum_{rs} \, \epsilon_{rs} \,c_r^{\dagger}\,c_s,  \label{E02}
\end{equation}
and\\
\begin{eqnarray}
\hat V_{2}& = &{\frac{1}{4}} \sum_{rstu} <rs|v_a|tu> 
                 c_r^{\dagger}c_s^{\dagger}c_u c_t \nonumber \\ 
     & = & \sum_{rstu \Gamma}\,{\frac{\sqrt{(2\Gamma+1)}}
                    {\sqrt{(1+\delta_{rs})(1+\delta_{tu})}}}  
                    \,<rs|v_a|tu>_{\Gamma} \nonumber\\
                    &&\qquad \qquad
                      \biggl( A_{\Gamma}^{\dagger }(rs)
                         \times \tilde A_{\Gamma}(tu)\biggr)_{0},
\label{E03}
\end{eqnarray}
where $\epsilon_{rs}$ are the single-particle energies 
of the spherical shell model
states, which are diagonal except in the radial quantum numbers and 
$<rs|v|tu>$ are the two-body interaction matrix elements and in the
present work are chosen to be those of ``USD''. The two-particle 
coupled operator in Eq. (3)
is given by $A_{\Gamma}^{\dagger }(rs)=
(c_{r}^{\dagger }c_{s}^{\dagger })_{\Gamma}$ and $%
\tilde A_{\Gamma M_\Gamma} = (-1)^{\Gamma - M_\Gamma} A_{\Gamma -M_\Gamma}$. The labels 
$r,s,...$ in the above equations denote the quantum numbers of angular-momentum
and isospin of the single-particle states.
``$\Gamma$'' quantum number labels both angular-momentum and isospin of the two-particle
coupled state. 
The above notation is same as that used in \cite{fhmw69}.

In the present work, the statistical averages have been calculated
using the canonical ensemble approach since the exact 
solutions have well defined
particle number. The average value of a physical quantity ``F'' in canonical
ensemble is given by \cite{fkms03,landau_stat}
\begin{equation}
\langle\langle F \rangle\rangle = {\sum_{i} F_i e^{-E_i/T}}/ {Z},
\end{equation}
where,
\begin{eqnarray}
Z     &=& \sum_{i} e^{-E_i/T}, \nonumber \\
\hat H |i\rangle &=& E_i |i\rangle, \nonumber \\
F_i   &=& \langle i|\hat F|i\rangle .
\end{eqnarray}
The statistical averages have been evaluated from the lowest 
one-thousand eigen-states, obtained from the diagonalization
of the shell model Hamiltonian, Eq. (1), for each angular-momentum. For a few
cases, we have also calculated the averages with 1500 eigen-states and results
were similar to those with one-thousand eigen-states.

The average quadrupole moment
of the shell model states has been calculated from the expectation value of the operator
\begin{equation}
\hat Q  = e_p \hat Q_p + e_n \hat Q_n
\end{equation}
where the quadrupole operator is given by
\begin{equation}
\hat Q_{p(n)} =  \sqrt{16\pi/5}\sum_{rs}
\langle r ||r^{2}Y_2 || s \rangle
(a^\dagger_r\times \tilde a_s)^2
\end{equation}
The effective charges $e_p=1.5$ and $e_n=0.5$ have been used and the harmonic oscillator
length parameter has been calculated with $\hbar \omega = 45 A^{-1/3} - 25 A^{-2/3}$ 
\cite{cwb86}.

The shell model calculations have been performed in the middle of the sd-shell for 
$^{28}$Si and $^{27}$Si with the USD interaction \cite{wild}.
The results for $^{28}$Si and  $^{27}$Si are
presented in Fig. 1 with the left side depicting the average 
quadrupole moment for  $^{28}$Si and the right side the results of $^{27}$Si. For  $^{28}$Si at low temperature,
the average quadrupole moment has a constant value of about $20~{\rm efm}^2$ up to 
temperature, T=1.2 MeV and above this temperature the quadrupole moment
drops. For higher temperatures, the quadrupole moment is noted
to approach  zero and, therefore, indicating that shape transition from
deformed to spherical shape has occurred. It is noted that the
drop in the quadrupole  moment is spread from T=1 to 4 MeV and it is 
not possible to determine the exact phase transitional point. In the grand
canonical mean-field calculations, the phase transition occurs at one point.
In the present canonical study, the phase transition is smeared out
due to the fluctuations present. The transitional point in quadrupole
moment can be approximately inferred as the middle of two points - one at which
the drop begins and the other point where the constant behavior is again observed.
These two temperature points for I=2 in Fig. 1 are 1 MeV and 4 MeV
and, therefore, the transitional point is inferred as T=2.5 MeV. 
It is to be noted
that positive quadrupole moment in Fig. 1 corresponds to the oblate shape for the ground-state
band of $^{28}$Si using the standard relationship between the laboratory and intrinsic quadrupole
moments \cite{cw88}.


For lighter mass sd-shell nuclei, the question of phase transition has been quite 
controversial. The mean-field HFB calculations predict the phase transition, evident 
from the peak structure obtained in the heat capacity \cite{hgmiller} at temperature, 
T=2.1 MeV and the vanishing of the quadrupole moment. The quadrupole moment
has a finite value at low temperatures and then depicts a transition at T=2.1 MeV.
The canonical shell model calculations, however, depicted a different behavior. The canonical heat capacity
does show a peak structure as in the grand-canonical mean-field
analysis, but the calculated quadrupole moments didn't vanish.
The analysis in these investigations
were performed mostly for $^{24}$Mg and all the angular-momentum states
were used to construct the canonical partition function \cite{bjcole,hg89}. 
In the present study of $^{28,27}$Si, the number of eigen-states for 
each angular-momentum are quite large and the partition function for
each angular-momentum can be independently constructed. 

It is evident from Fig. 1 that the quadrupole-moments show a
transitional behavior for each calculated angular-momentum ensemble. The interesting 
observation from Fig. 1 is that the drop in the quadrupole-moment is shifted to lower 
temperatures with increasing angular-momentum. For I=4, the drop in quadrupole-moment
is noted to start at T=0.8 MeV and the transitional point is inferred to be
T=2.3 MeV. For I=6, the drop in the quadrupole moment occurs at a slightly lower 
temperature as compared to I=4 and for I=8 the drop starts at a very low temperature.
We have also analyzed the temperature behavior of angular-momentum,
I=1,3, 5 and 7 ensembles and for these ensembles the quadrupole moments depict a very irregular dependence on 
temperature. This would explain the reason that the quadrupole moments in the
earlier study didn't depict a transitional behavior as these quantities were averaged
over all angular-momentum ensembles.
Further, results of the present work are in agreement with the mean-field results
which predict the transitional behavior for both quadrupole-moment as well
as the heat capacity \cite{hg89}. Although, these mean-field calculations were performed
for $^{24}$Mg, but it is expected that these results should be similar for $^{28}$Si.
 
The results for $^{27}$Si, presented on the right side in  Fig. 1, for the angular-momentum 
I = 5/2, 9/2, 13/2 and 17/2 ensembles, show that the drop in the quadrupole moment has 
a similar behavior as that of the even-even system, $^{28}$Si.
The transitional point in the quadrupole moment is inferred to be 
approximately T=2.25 MeV for I=5/2 . For I=9/2, this transitional point
is slightly lowered and occurs at T=2 MeV. For I=13/2 and 17/2, the transition
also occurs at about T=2 MeV.

In Fig.~1, quadrupole deformations have been plotted as a function of
temperature for a fixed value of angular-momentum.
It is also quite instructive
to show the angular-momentum dependence of the quadrupole moment for a fixed
value of temperature. In Fig. 2, angular-momentum dependence of the quadrupole deformation for 
temperatures of 0, 1, 2 and 3 MeV is plotted. It is observed from the figure that as the 
temperature is raised, the quadrupole deformation of higher angular momentum states 
tends to drop. A clear trend of quadrupole moments dropping with increasing temperature 
for larger angular-momentum ensembles is evident from Fig. 2.


The advantage in the shell model study is that the analysis can be performed for each
individual state \cite{zel}. In Fig. 3, the quadrupole moments are shown for each 
individual eigen-state and have been plotted for the first one-thousand states for both 
the studied nuclei. The quadrupole moment for the low-lying states are randomly 
distributed from large positive to negative deformations. However, with
increasing eigen-state number, it is evident that the quadrupole moments converge to
the spherical shape.
In particular, for I=2 and 5/2, most of the states at higher excitation 
energy have zero quadrupole moment. The fluctuations from zero quadrupole moment increases 
with increasing spin and is due to the reason that the dimensionality
of the basis states becomes progressively smaller with spin.


In conclusion, temperature and angular momentum dependence of the quadrupole deformation has been
studied in the middle of the sd-shell. It is quite evident from the
present study that collective states for the studied nuclei, $^{28}$Si and 
$^{27}$Si, depict a transitional behavior from large oblate deformation to spherical shape. 
This transition is evident from both canonical ensemble study and the 
quadrupole moments of the individual states.

Finally, we would like to mention that the results of the present work are questionable due to restricted 
sd-shell configuration space for T $\gtrsim$ 2.5 MeV. For higher temperatures, it is expected that the fp-shell will be populated. 
However, the inclusion of fp shell configuration space is almost impossible in the present context as one needs to 
evaluate at least 1000 to 1500 eigen-states for each angular-momentum to have a proper statistical description. The 
shell model Monte Carlo (SMMC) approach \cite{lang95} is a possible solution to include the fp-shell and we intend to look 
into this problem in the near future. 


\newpage 
\begin{figure}[]
\epsfxsize=12cm
\centerline{\epsfbox{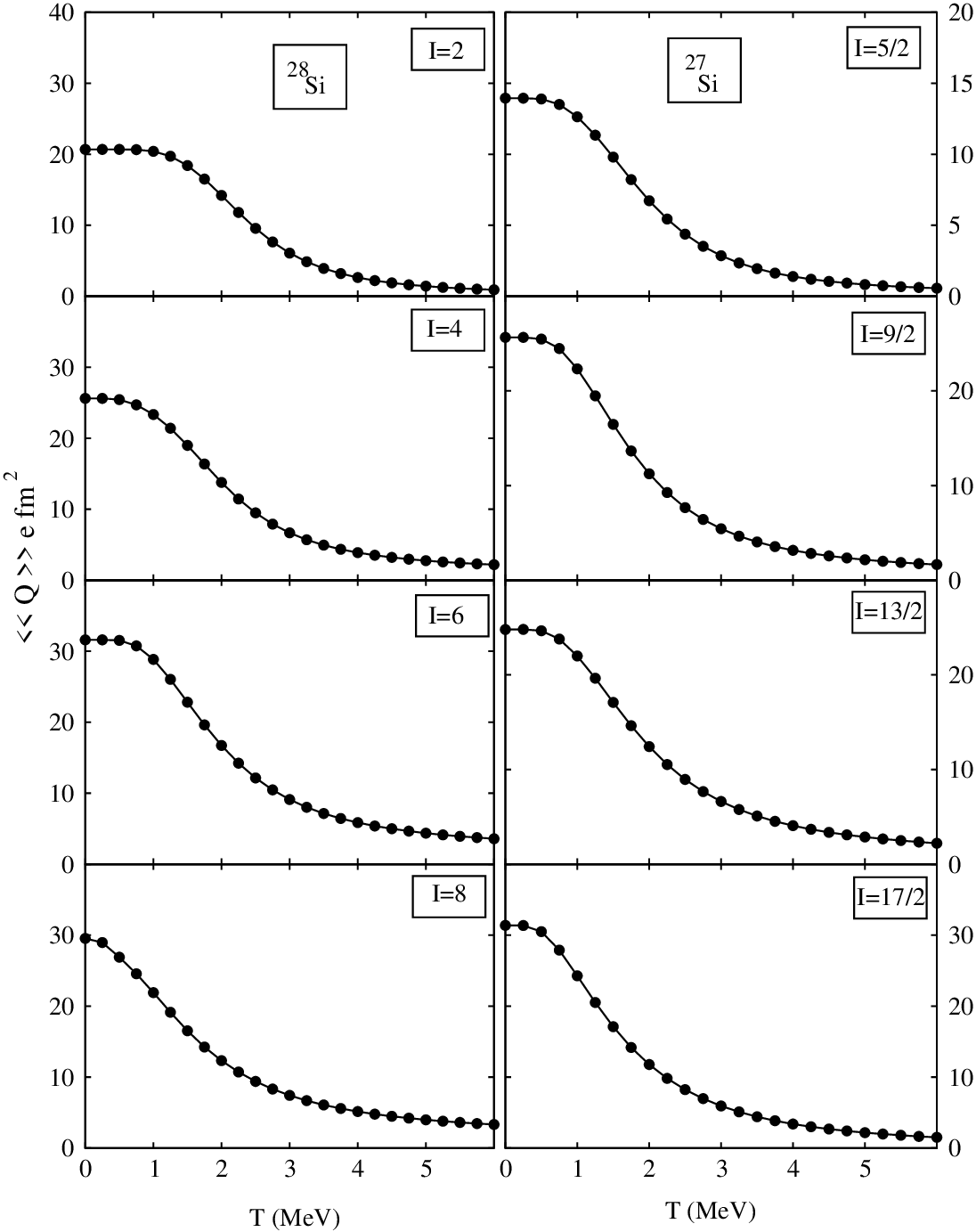}}
\caption{Quadrupole moments are plotted as a function of temperature  
for various angular-momentum states. The left panel depicts the results for  
$^{28}$Si and the right panel for  $^{27}$Si.}
\label{figure.1}
\end{figure}

\begin{figure}[]
\epsfxsize=12cm
\centerline{\epsfbox{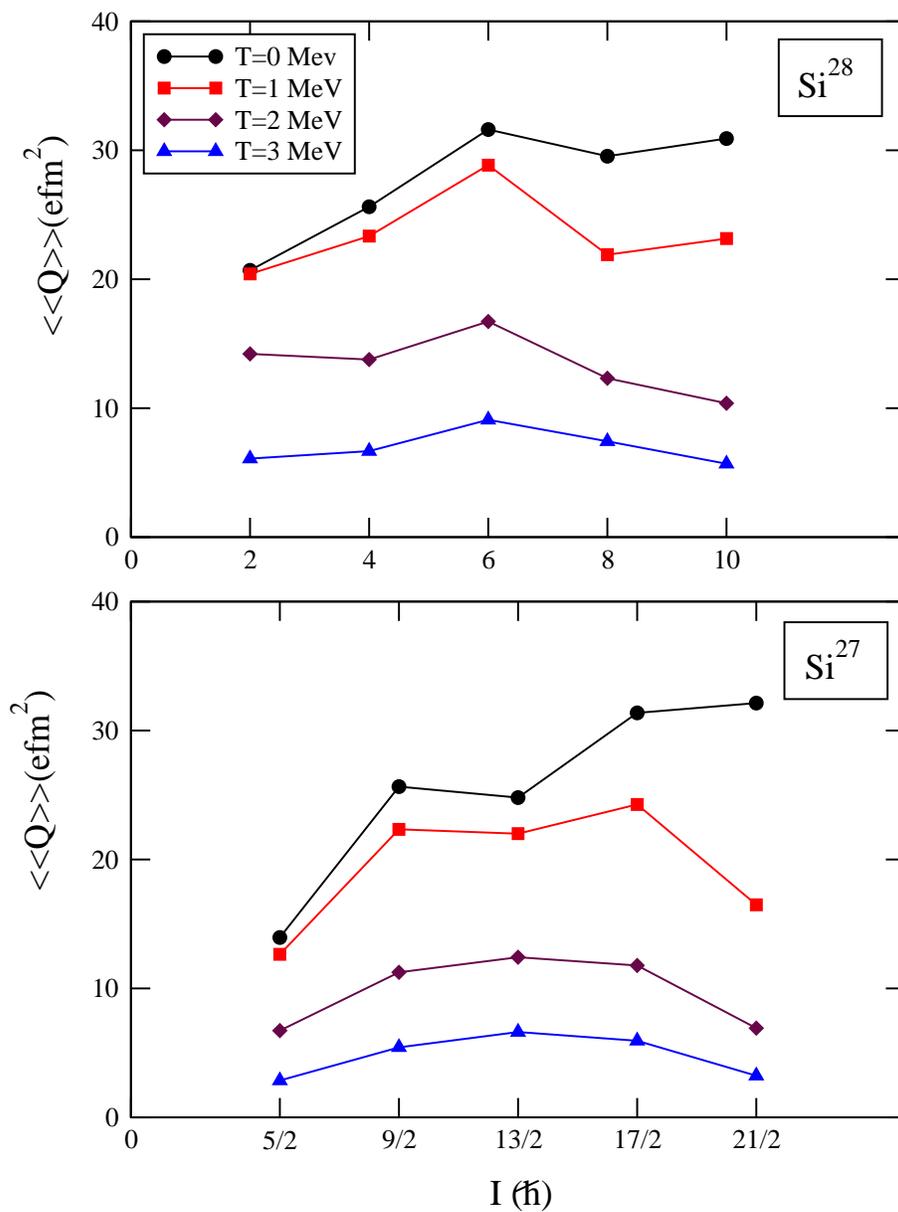}}
\caption{ Angular momentum dependence of the quadrupole deformation for temperatures of 0, 1, 2 and 3 MeV 
is plotted  for the two studied isotopes of $^{28}$Si and $^{27}$Si.}
\label{figure.2}
\end{figure}

\begin{figure}[]
\epsfxsize=12cm
\centerline{\epsfbox{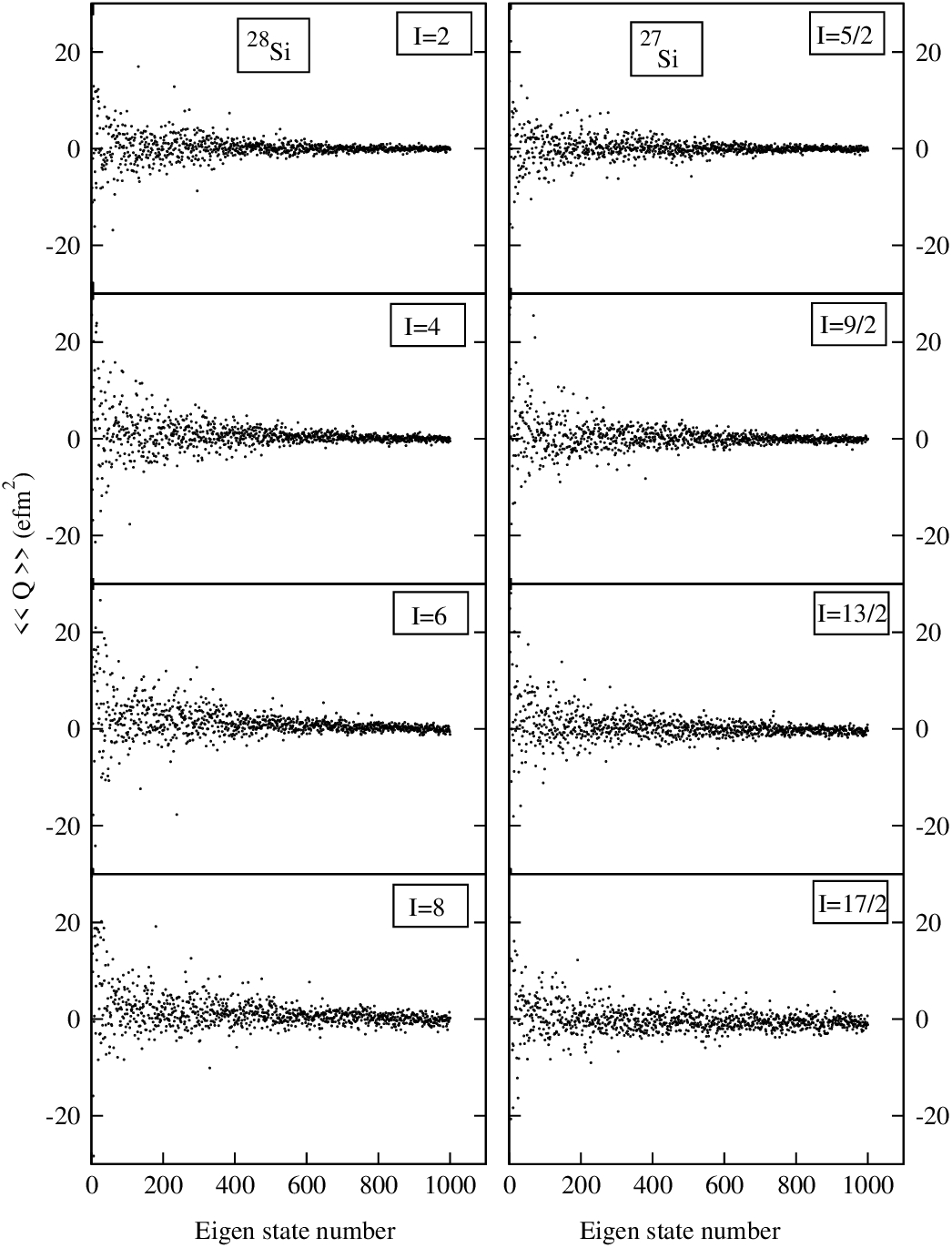}}
\caption{ Quadrupole moments are shown for the first one-thousand individual
eigen-states of the two studied isotopes of $^{28}$Si and $^{27}$Si.
}
\label{figure.3}
\end{figure}

\end{document}